# Giant Flexoelectric Effect in Ferroelectric Epitaxial Thin Films

Daesu Lee,[1] A. Yoon,[2] S. Y. Jang,[1] J.-G. Yoon,[3] J.-S. Chung,[4] M. Kim,[2] J. F. Scott,[5] and T. W. Noh[1,*]

[1] *ReCFI, Department of Physics and Astronomy, Seoul National University, Seoul 151-747, Korea*

[2] *Department of Materials Science and Engineering, Seoul National University, Seoul 151-747, Korea*

[3] *Department of Physics, University of Suwon, Suwon, Gyunggi-do 445-743, Korea*

[4] *Department of Physics and CAMDRC, Soongsil University, Seoul 156-743, Korea*

[5] *Department of Physics, University of Cambridge, Cambridge CB3 0HE, UK*

We report on nanoscale strain gradients in ferroelectric $HoMnO_3$ epitaxial thin films, resulting in a giant flexoelectric effect. Using grazing-incidence in-plane X-ray diffraction, we measured strain gradients in the films, which were 6 or 7 orders of magnitude larger than typical values reported for bulk oxides. The combination of transmission electron microscopy, electrical measurements, and electrostatic calculations showed that flexoelectricity provides a means of tuning the physical properties of ferroelectric epitaxial thin films, such as domain configurations and hysteresis curves.

*Author to whom correspondence should be addressed: twnoh@snu.ac.kr

The flexoelectric effect describes an electric field that is generated by a strain gradient, and vice versa, whereas conventional electromechanical couplings such as piezoelectricity generally assume homogeneous strain conditions [1–5]. Because strain gradients break inversion symmetry, flexoelectricity allows the generation of electric responses from lattice deformations in every dielectric material; it occurs in all 32 crystalline point groups (unlike piezoelectricity, which exists only in noncentrosymmetric systems of 20 point groups). Owing to this universal nature, flexoelectricity has inspired a wide range of scientific interest and has broad application potential. Particularly in flexible systems such as liquid crystals [6], low-dimensional crystals (e.g., graphene or carbon nanotubes) [7], and biological molecular membranes or hairs [8], the flexoelectric effect can be quite significant.

In solids, however, there has been little investigation into the flexoelectric effect. One of the major reasons for this lack of research is the minuscule magnitude of the effect. The flexoelectric coefficient $f$ is quite small (i.e., $f \sim e/a$, where $e$ is the electronic charge and $a$ is the lattice constant), and elastic deformation in most solids is limited [9]. Additionally, it is difficult to achieve adequate control of the strain gradient through the application of external stress. Thus, the basic issue of whether flexoelectricity emerges as a practical means to tune physical properties in solids is untouched to date.

In this Letter, we develop a general framework for realizing and modulating the giant flexoelectric effect in epitaxial oxide thin films, emphasizing the key role of flexoelectricity in solids. In epitaxial oxide thin films, a lattice mismatch between the film and the substrate can result in strain relaxation within tens of nanometers of the film/substrate interface, inducing a large strain gradient. We used tensile-strained $HoMnO_3$ epitaxial thin films as a model system [10–12]. $HoMnO_3$ is a ferroelectric material below $T_C \approx 875$ K with a moderate remnant ferroelectric polarization ($P \approx 5.6$ μC cm$^{-2}$) [13,14]. Its ferroelectric characteristics can be tuned by the flexoelectricity-induced electric field; thus, this material offers a good opportunity to investigate the flexoelectric effect in epitaxial ferroelectric films.

We begin with a structural analysis of the films deposited at different oxygen partial pressures, $P_{O2}$, using high-resolution X-ray diffraction (XRD). $HoMnO_3$ films deposited at $P_{O2}$ = 10 mTorr and $P_{O2}$ = 350 mTorr are abbreviated as HMO10 and HMO350, respectively. Figure 1(a) shows that the out-of-plane lattice constant $c$ increased with decreasing $P_{O2}$. The out-of-plane lattice constant in HMO350 ($c$ = 11.337 Å) was smaller than the bulk constant ($c_{bulk}$ = 11.406 Å). By contrast, the out-of-plane lattice constant in HMO10 ($c$ = 11.575 Å) was larger than the bulk constant, although these films experienced in-plane tensile strain. The increase in crystal volume at low $P_{O2}$ can be attributed to oxygen vacancies in the oxides [15].

The XRD patterns in Figure 1(a) show a shoulder at the right-hand side of the (0004) diffraction peak. For HMO350, we observed a clear shoulder on the right-hand side, which is indicative of a tensile-strain gradient [16]. The shoulder feature weakened with decreasing $P_{O2}$. The XRD $\theta$–$2\theta$ data indicate that the average tensile-strain gradient was reduced because of increasing crystal volume under low-$P_{O2}$ conditions (Figs. 1(b) and 1(c); see Supplemental Material [10]).

Further information on the strain gradient in epitaxial films can be obtained more directly from high-resolution, grazing-incidence in-plane XRD (GIXRD). GIXRD is a powerful tool for determining the depth profile of the in-plane lattice constant $a$, from which the strain and the strain gradient can be estimated [10]. Our GIXRD measurements were performed using a six-circle XRD machine using synchrotron radiation. The penetration depth $L$ of the X-rays is proportional to the grazing-incidence angle $\lambda$. While the lattice constant averaged over the entire film region can be measured with a large $\lambda$, we obtained information on the value of $a$ near the film surface with a small $\lambda$. Figure 2(a) shows that the averaged in-plane tensile strain $\bar{u}$ = ($a_{film}$ − $a_{bulk}$)/$a_{bulk}$ decreased exponentially with decreasing $L$ [17]. The magnitude of the decrease was larger and more abrupt in HMO350 than HMO10, as expected based on Figures 1(b) and 1(c).

The strain gradient $\partial u/\partial z$ estimated from the GIXRD data as a function of the distance $z$ from the film surface was as large as $10^5$–$10^6$ m$^{-1}$ (Fig. 2(b)) [10]. Note that these $\partial u/\partial z$ values in the HoMnO$_3$ epitaxial thin films are 6 or 7 orders of magnitude larger than previous values reported for bulk solids, which are on the order of 0.1 m$^{-1}$ [9]. This giant strain gradient can occur in thin-film epitaxy because of the large lattice mismatch between the film and the substrate (+3.5% with Pt(111)/Al$_2$O$_3$(0006) substrates), which leads to strain relaxation within a few tens of monolayers of the interface [18]. If we assume that 1% of the strain relaxes within 10 nm of the film/substrate interface, the strain gradient becomes $10^6$ m$^{-1}$, which is similar to our reported value. Furthermore, the strain gradient can be modulated by varying $P_{O2}$ during the growth process. Because of the large $\partial u/\partial z$, which can also be modulated, the flexoelectric effect becomes significant in epitaxial oxide thin films.

The strain gradient generates an internal electric field $E_S$ due to the flexoelectric effect [3–5,19], which can be expressed as follows:

$$E_S = \frac{e}{4\pi\varepsilon_0 a}\frac{\partial u}{\partial z}, \qquad (1)$$

where $e$ is the electronic charge, $\varepsilon_0$ is the permittivity of free space, and $\partial u/\partial z$ is the strain gradient. We estimated $E_S$ in our samples by inserting the experimental values of $\partial u/\partial z$ into Eq. (1). We found that $E_S = 0.7$ MV m$^{-1}$ (i.e., $7\times10^{-4}$ V nm$^{-1}$) in HMO10 and $E_S = 5.0$ MV m$^{-1}$ (i.e., $5\times10^{-3}$ V nm$^{-1}$) in HMO350 at room temperature. The estimated value of $E_S$ seems small compared with the room-temperature ferroelectric coercive field, which was ~40 MV m$^{-1}$ in our HoMnO$_3$ films. However, the coercive field becomes much smaller at temperatures close to $T_C$, where the ferroelectric interaction is weaker [20]. On the other hand, the temperature dependence of $E_S$ is expected to be weak because the HoMnO$_3$ film and substrate have similar thermal-expansion coefficients [21,22]. Thus, near $T_C$, $E_S$ can be comparable to the coercive field, and the associated flexoelectric effect can become significant.

As shown schematically in Figure 3(a), $E_S$ played an important role in determining the domain configurations at temperatures near $T_C$. The HoMnO$_3$ films were deposited at 860°C, substantially higher than $T_C$, and cooled slowly [10]. For the film deposited at low $P_{O2}$, the strain gradient was low, and a mixed polydomain formed in the film. Polydomain formation is typical in ferroelectrics because it reduces the depolarization energy. For the film deposited at high $P_{O2}$, however, a large strain gradient occurred. Under high-$P_{O2}$ deposition, $E_s$ can be large enough that a monodomain forms at temperatures near $T_C$. The effect of $E_S$ on the domain configuration near $T_C$ was further investigated by performing quantitative electrostatic calculations as a function of $E_S$ and temperature. Our calculations (see Supplemental Material [10]) showed that as $E_S$ increased, the domain configurations changed from polydomain to monodomain. Additionally, larger $E_S$ provided a wider temperature window for monodomain formation.

Interestingly, the tuning parameter during the growth process was simply $P_{O2}$. We observed a large variation in domain configurations in our HoMnO$_3$ thin films as a function of $P_{O2}$. Figures 4(a) and 4(c) show the ferroelectric domains in our films, measured using transmission electron microscopy (TEM) with dark-field imaging. Ferroelectric domains can be imaged by dark-field TEM due to failure of Friedel's law in noncentrosymmetric structures [23]. The bright and dark regions in the TEM images correspond to up and down domains, respectively. From the intensity profiles (Figs. 4(b) and 4(d)), we measured the width of the domains, denoted by $\alpha$ and $\beta$ for up and down domains, respectively. We evaluated averaged ratios of up- to down-domain widths and found that $\overline{\alpha}/\overline{\beta} = 0.8$ for $P_{O2} = 20$ mTorr and $\overline{\alpha}/\overline{\beta} = 3.2$ for $P_{O2} = 300$ mTorr. As $P_{O2}$ increased, the domain configuration changed from an up/down mixed pattern to one in which up domains are preferred.

Domain configurations determined by the flexoelectric effect are more important in the presence of irreversible defect dipoles [20]. The presence of defect dipoles was reported previously in our films [12] and other ferroelectric films [24]. Figure 3(b) shows a schematic diagram that explains how the $P_{O2}$-modulated flexoelectric effect can affect the alignment of defect dipoles. These defect dipoles align parallel to the ferroelectric polarization $P_{FE}$ because of the site preference of point defects [12,24,25]. Namely, the alignment of defect dipoles in our HoMnO$_3$ films can differ according to the domain configurations. The alignment of defect dipoles is generally preserved at room temperature; thus, domains become pinned according to the high-temperature alignment of the defect dipoles. Domain switching can be influenced by domain pinning, which induces a modification in the polarization–electric field (P–E) hysteresis loops (Fig. 3(c)) [20,26].

We found that the P–E hysteresis loops were largely dependent on $P_{O2}$. Figure 5(a) shows that as $P_{O2}$ increased, the P–E curve changed from a double loop at 10 mTorr to an asymmetric double loop at 100 mTorr and to a nearly biased single loop at 350 mTorr. This variation in the P–E hysteresis loops is consistent with our prediction above (see Fig. 3(c)). Additionally, the systematic change in the P–E loops is confirmed by the curve of the switching current $I_{SW}$ (Fig. 5(b)), where its time integration yields P [27]. Because bulk HoMnO$_3$ exhibits simple symmetric P–E hysteresis loops below $T_C = 875$ K [14], the large systematic modification in the P–E loops that we observed is a novel phenomenon. These results demonstrate that we can modify the P–E hysteresis loops simply by varying $P_{O2}$.

This mechanism of giant flexoelectric effects provides general insight into thin-film epitaxy. Strain due to the lattice mismatch between the substrate and epilayer is the typical parameter used to tune the physical properties of thin films [11,28]. However, the strain gradient must also be considered in any epitaxial system in which strain relaxation occurs. Specifically, as we demonstrated, the effects of the strain gradient can be significant near $T_C$ in ferroelectric epitaxial thin films, such as HoMnO$_3$ [12], BiFeO$_3$ [24], and highly strained ferroelectric films [28]. Our proposed mechanism provides an explanation for the deformation of *P–E* hysteresis loops commonly observed in some ferroelectric epitaxial films.

In summary, we demonstrated that the flexoelectric effect in ferroelectric epitaxial films can be extremely large, and furthermore, can be modulated. We have shown that the flexoelectric effect can strongly affect polarization hysteresis curves as well as domain configurations. Because the strain gradient can be generated readily in epitaxial oxide thin films via strain relaxation, our findings can be used to tune the physical properties of films using flexoelectric and/or flexomagnetic effects.

The authors thank Professor G. Catalan and Dr. P. Zubko for valuable discussions. This research was supported by a National Research Foundation of Korea (NRF) grant funded by the Korean Government (MEST, No. 2010-0020416). X-ray measurements were performed at the 10C1 beamline of Pohang Light Source. D. L. acknowledges support from the POSCO TJ Park Doctoral Foundation.

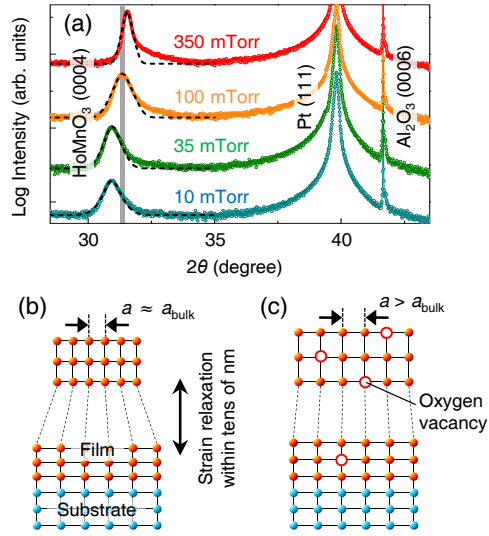

FIG. 1 (color online). (a) XRD $\theta$–$2\theta$ scans of HoMnO$_3$ films. The gray solid vertical line indicates the position of the (0004) diffraction peak of bulk HoMnO$_3$. Schematic diagrams show tensile-strain relaxation (b) for typical growth conditions and (c) for a film with more oxygen vacancies, which induced crystal-volume expansion. Because of various relaxation mechanisms, we represent the strain-relaxation process by dashed guidelines.

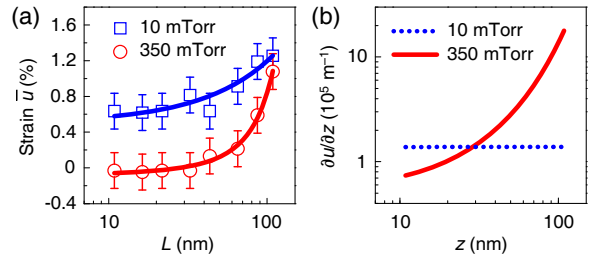

FIG. 2 (color online). (a) Variation in the averaged in-plane strain $\bar{u}$ as a function of penetration depth $L$ in HMO10 and HMO350. The solid curves are fitted results. (b) Estimated strain gradient in HMO10 and HMO350 as a function of the distance from the film surface $z$.

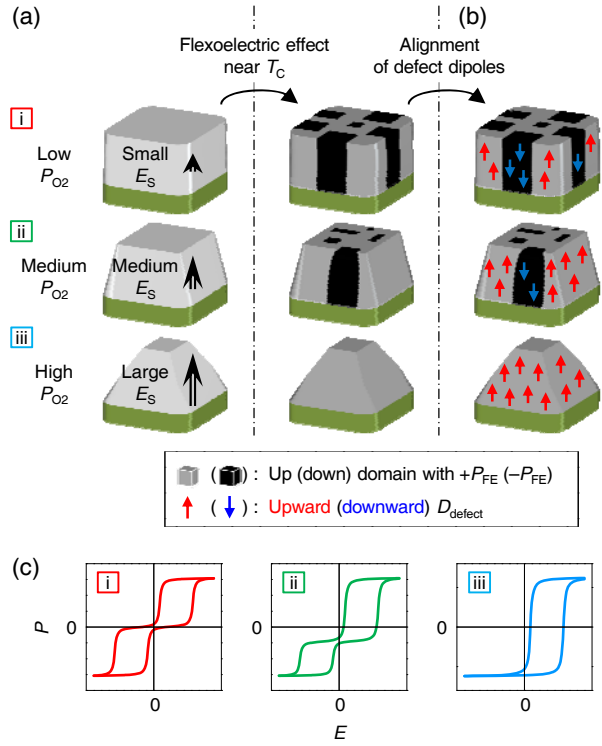

FIG. 3 (color online). (a) Schematic diagram showing how $E_S$ can affect domain configuration, which depends on $P_{O2}$ during the growth process. (b) At high growth temperatures, defect dipoles can align along the polarization of each domain. (c) Expected variations in the *P–E* hysteresis loops according to the alignment of the defect dipoles, which can depend on $P_{O2}$.

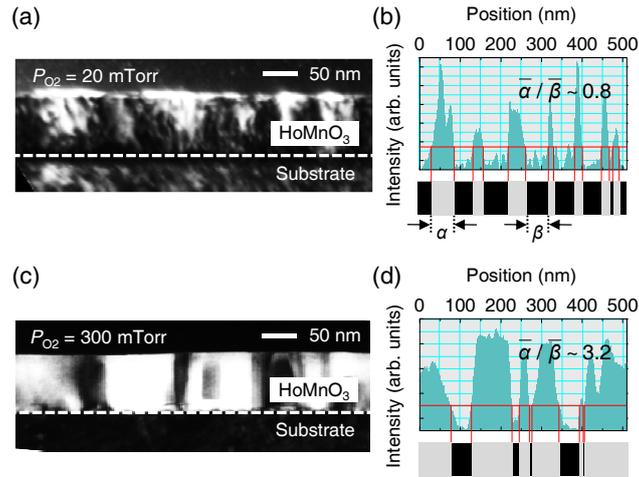

FIG. 4 (color online). Ferroelectric domain images, measured using dark-field TEM imaging, of the as-grown film deposited at a $P_{O2}$ of (a) 20 mTorr and (c) 300 mTorr. The bright and dark regions are attributed to the up- and down-polarization domains, respectively. (b) and (d) show intensity profiles of the TEM images, from which the averaged width $\bar{\alpha}$ ($\bar{\beta}$) of the up-polarization (down-polarization) domains were estimated.

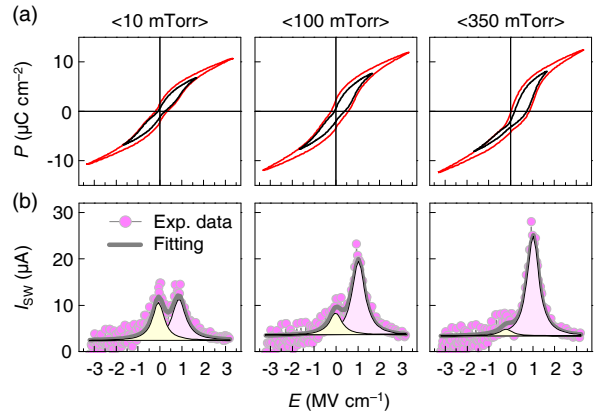

FIG. 5 (color online). Systematic variation in (a) the $P$–$E$ and (b) $I_{SW}$ hysteresis loops from double loops to biased single loops for different $P_{O2}$ during the growth process. $I_{SW}$ was measured during the polarization-switching process from down to up domains, and the $I_{SW}$ data were well fitted by the sum (gray solid line) of two Lorentzian curves (black solid lines).

# Supplemental Material for

# "Giant Flexoelectric Effect in Ferroelectric Epitaxial Thin Films"

D. Lee,[1] A. Yoon,[2] S. Y. Jang,[1] J.-G. Yoon,[3] J.-S. Chung,[4] M. Kim,[2] J. F. Scott,[5]

and T. W. Noh[1, *]

[1] *ReCFI, Department of Physics and Astronomy, Seoul National University,*

*Seoul 151-747, Korea*

[2] *Department of Materials Science and Engineering, Seoul National University, Seoul 151-*

*747, Korea*

[3] *Department of Physics, University of Suwon, Suwon, Gyunggi-do 445-743, Korea*

[4] *Department of Physics and CAMDRC, Soongsil University, Seoul 156-743, Korea*

[5] *Department of Physics, University of Cambridge, Cambridge CB3 0HE, UK*

---

[*] E-mail: twnoh@snu.ac.kr

**Methods**

HoMnO$_3$ epitaxial films (~100 nm thick) were grown on Pt(111)/Al$_2$O$_3$(0006) and yttria-stabilized zirconia (YSZ) (111) substrates via pulsed laser deposition at 860°C using a KrF excimer laser (wavelength 248 nm; Lambda Physik) [1]. HoMnO$_3$ has a lattice mismatch of ($a_{\text{substrate}} - a_{\text{HMO}})/a_{\text{HMO}}$) = 3.5% with Pt(111)/Al$_2$O$_3$(0006), and a lattice mismatch of ($a_{\text{substrate}} - a_{\text{HMO}})/a_{\text{HMO}}$) = 2.8% with YSZ(111). We varied $P_{\text{O2}}$ during growth from 10 to 350 mTorr. XRD $\theta$–$2\theta$ measurements were performed at room temperature using a high-resolution four-circle X-ray diffractometer (D8 Advance, Bruker AXS). For electrical measurements, 80-μm-diameter gold-top electrodes were deposited using a shadow mask. We measured the $P$–$E$ and $I_{\text{SW}}$ hysteresis loops of the films using a low-temperature probe station (Desert Cryogenics) and TF Analyzer (aixACCT) at temperatures in the range of 10–200 K and at a frequency of 2 kHz. Figure 5 shows data measured at 100 K. We carried out TEM measurements using dark-field imaging for the HoMnO$_3$-on-YSZ films. To prepare a high-quality TEM sample, we selected a YSZ substrate with a clean top surface and a lattice mismatch similar to the Pt/Al$_2$O$_3$ substrate. Samples for cross-sectional TEM imaging were prepared using the conventional method with mechanical thinning followed by Ar$^+$ milling at 3 kV; the transmission electron microscope (JEOL-2010F) was operated at 200 kV. For two-beam dark-field images, the sample was tilted from the [010] zone axis by ~3° to excite the 300 spot.

**Variation in strain gradient by $P_{O2}$ growth conditions**

According to a general model for the strain profile [2,3], independent of the actual relaxation mechanism, the strain $u(z)$ in epitaxial thin films can be expressed as follows:

$$u(z) = u_0 + u_1 \cdot e^{-\alpha(t-z)}, \qquad (S1)$$

where $u_0$, $u_1$, and $\alpha$ are constants, $t$ is the thickness of the film, and $z$ is the distance from the film surface. Thus, Eq. (S1) enables a rough estimation of the strain gradient if the two end values of $u$ are known, i.e., the $u$ values at the interface ($z = t$) and at the surface ($z = 0$). Typically, a film can be fully strained near the interface, and thus, the $u$ value at the interface ($z = t$) can be determined by the misfit strain relative to the substrate (note that in our study, only the tensile misfit strain (i.e., $u \geq 0$) was considered). Near the surface, the misfit strain is usually relaxed, and the $u$ value at the surface ($z = 0$) can converge to zero for sufficiently thick films. That is, the lattice constant $a$ relaxes from $a_{substrate}$ to $a_{bulk}$ (Fig. S1). However, if crystal-volume expansion occurs due to oxygen vacancies ($V_O$), the $u$ value at the surface ($z = 0$) can converge to a nonzero positive value. That is, $a$ relaxes from $a_{substrate}$ to an expanded quantity $a_{expanded}$ ($> a_{bulk}$) (Fig. S1). The amount of $V_O$ in films can be adjusted by varying $P_{O2}$ during film growth. Thus, using the simple model equation (Eq. S1) of strain relaxation and crystal-volume expansion, we can explain how the strain gradient can be modulated by varying $P_{O2}$.

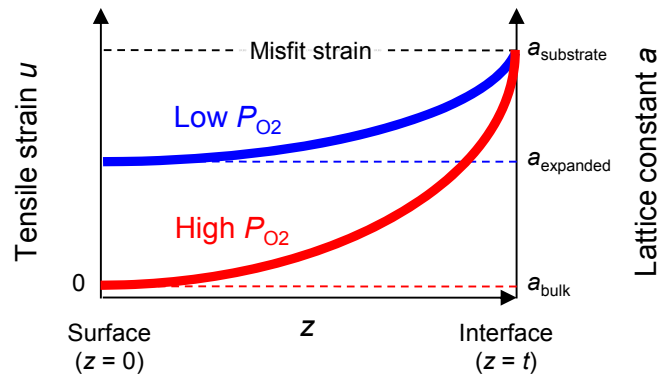

**Figure S1.** Schematic description of strain relaxation according to $P_{O2}$.

**Grazing-incidence in-plane XRD (GIXRD) measurements**

GIXRD is a powerful tool for determining the depth profile of the in-plane lattice constant $a$, from which the strain and the strain gradient can be estimated [4]. We determined the in-plane strain by directly measuring the spacing among the HoMnO$_3$ ($11\bar{2}0$) Bragg planes, as schematically shown in Figure S2. Our GIXRD measurements were performed using a six-circle XRD instrument with synchrotron radiation at the Pohang Light Source (PLS) on beamline 10C1. The penetration depth of the X-rays is proportional to the grazing-incidence angle $\lambda_i$. While the lattice constant averaged over the entire film region can be measured with a large $\lambda_i$, we obtained information on the value of $a$ near the film surface with a small $\lambda_i$.

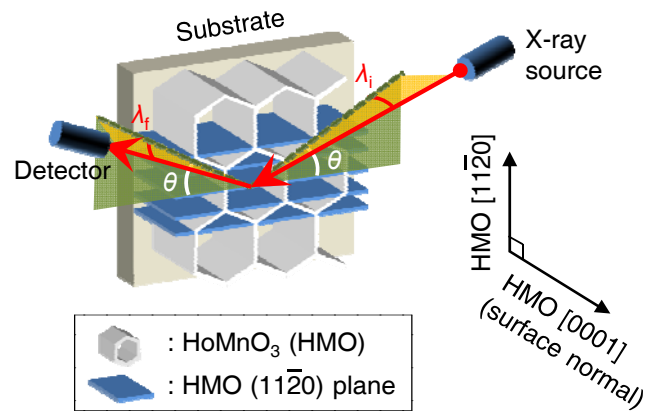

**Figure S2.** GIXRD using a six-circle X-ray diffractometer.

**Estimation of strain gradient**

From Eq. (S1), we obtained the averaged strain $\overline{u(z)}$:

$$\overline{u(z)} = \frac{\int_0^z u(x)\mathrm{d}x}{z} = u_0 + \left(\frac{u_1 e^{-\alpha t}}{\alpha}\right)\left(\frac{e^{\alpha z}-1}{z}\right). \tag{S2}$$

Using Eq. (S2), we fitted the experimental data (open symbols in Fig. 2(a)), and determined the functional form of $\overline{u(z)}$. Fits to $\overline{u(z)}$ are shown as solid lines in Figure 2(a). Finally, we estimated $\partial u(z)/\partial z$ from $\overline{u(z)}$ by following the procedure shown below:

$$\begin{aligned}
\overline{u(z)} &= \frac{\int_0^z u(x)\mathrm{d}x}{z} \rightarrow z \cdot \overline{u(z)} = \int_0^z u(x)\mathrm{d}x \\
&\rightarrow u(z) = \overline{u(z)} + z \cdot \frac{\partial \overline{u(z)}}{\partial z} \rightarrow \frac{\partial u(z)}{\partial z} = 2\frac{\partial \overline{u(z)}}{\partial z} + z \cdot \frac{\partial^2 \overline{u(z)}}{\partial z^2}
\end{aligned}. \tag{S3}$$

**Electrostatic calculations**

To investigate the effect of $E_S$ on the domain configurations at temperatures near $T_C$, we performed electrostatic calculations as a function of $E_S$ and temperature. For simplicity, we assumed two-dimensional periodic square-like 180° domains (Fig. S3).

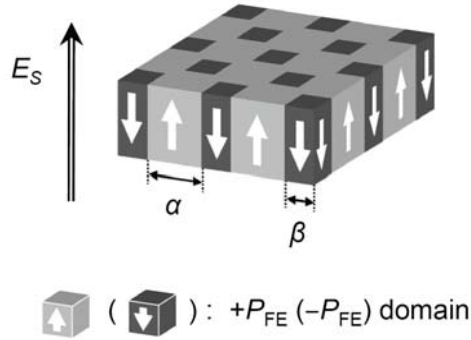

**Figure S3.** A simple model of two-dimensional periodic square-like 180° domains. $\alpha$ and $\beta$ indicate the width of the up and down domains, respectively.

First, in this model, we found the Fourier expansion of $\vec{P}(x,y)$ as a function of the polarization, $P_{FE}$, and the domain widths, $\alpha$ and $\beta$ as follows:

$$P_z(x,y) = \frac{P_{FE}}{(\alpha+\beta)^2} \cdot [(\alpha+\beta)^2 - 4\beta^2] + \frac{4\beta P_{FE}}{(\alpha+\beta)\pi} \cdot [\sum_{n=1}^{\infty}\{\frac{1}{n}(\sin(\frac{n\pi\alpha}{\alpha+\beta}) - \sin(\frac{n\pi\beta}{\alpha+\beta}))\cos(\frac{2n\pi}{\alpha+\beta}x)\}$$
$$+ \sum_{m=1}^{\infty}\{\frac{1}{m}(\sin(\frac{m\pi\alpha}{\alpha+\beta}) - \sin(\frac{m\pi\beta}{\alpha+\beta}))\cos(\frac{2m\pi}{\alpha+\beta}y)\}]$$
$$+ \frac{8P_{FE}}{\pi^2} \cdot \sum_{n,m=1}^{\infty}\{\frac{1}{nm}\cdot(\sin(\frac{n\pi\beta}{\alpha+\beta})\sin(\frac{m\pi\alpha}{\alpha+\beta}) + \sin(\frac{n\pi\alpha}{\alpha+\beta})\sin(\frac{m\pi\beta}{\alpha+\beta}))\cos(\frac{2n\pi}{\alpha+\beta}x)\cos(\frac{2m\pi}{\alpha+\beta}y)\}$$

(S4)

$$P_x(x,y) = P_y(x,y) = 0.$$  (S5)

Figure S4 shows an example of the Fourier expansion of $P_z(x, y)$ for $\alpha = 3$ and $\beta = 1$.

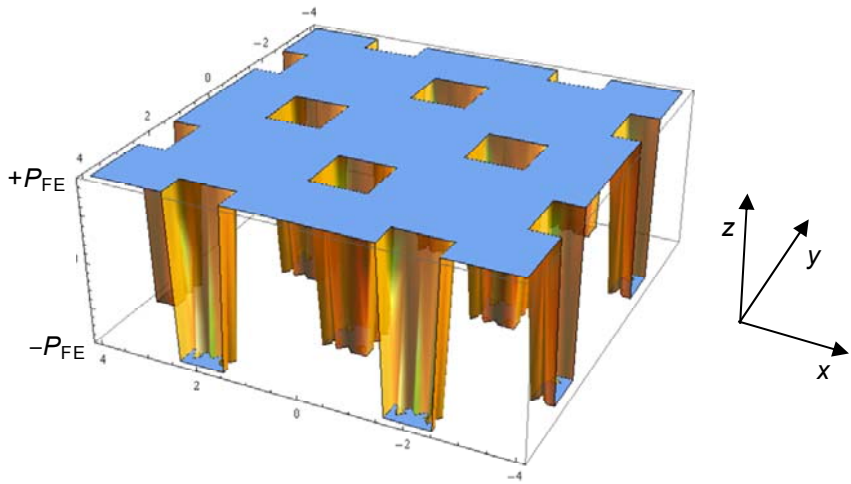

**Figure S4.** Fourier expansion of $P_z(x, y, z)$ for the two-dimensional periodic 180° domain with $\alpha = 3$ and $\beta = 1$.

We obtained the electric displacement field, $\vec{D}(x,y,z)$, by exploiting the symmetry of the system and the fact that $\nabla \cdot \vec{D} = \rho_f = 0$ and $\nabla^2 \vec{D} = \nabla^2 \vec{P}$, as shown below:

$$D_z(x,y,z) = \frac{P_{FE}}{(\alpha+\beta)^2} \cdot [(\alpha+\beta)^2 - 4\beta^2] + \frac{4\beta P_{FE}}{(\alpha+\beta)\pi} \cdot [\sum_{n=1}^{\infty} \{\frac{1}{n}(\sin(\frac{n\pi\alpha}{\alpha+\beta}) - \sin(\frac{n\pi\beta}{\alpha+\beta}))\cos(\frac{2n\pi}{\alpha+\beta}x) \cdot e^{\frac{-2n\pi}{\alpha+\beta}z}\}$$

$$+ \sum_{m=1}^{\infty} \{\frac{1}{m}(\sin(\frac{m\pi\alpha}{\alpha+\beta}) - \sin(\frac{m\pi\beta}{\alpha+\beta}))\cos(\frac{2m\pi}{\alpha+\beta}y) \cdot e^{\frac{-2m\pi}{\alpha+\beta}z}\}]$$

$$+ \frac{8P_{FE}}{\pi^2} \cdot \sum_{n,m=1}^{\infty} \{\frac{1}{nm} \cdot (\sin(\frac{n\pi\beta}{\alpha+\beta})\sin(\frac{m\pi\alpha}{\alpha+\beta}) + \sin(\frac{n\pi\alpha}{\alpha+\beta})\sin(\frac{m\pi\beta}{\alpha+\beta}))\cos(\frac{2n\pi}{\alpha+\beta}x)\cos(\frac{2m\pi}{\alpha+\beta}y) \cdot e^{\frac{-2\sqrt{n^2+m^2}\pi}{\alpha+\beta}z}\}$$

, (S6)

$$D_x(x,y,z) = \frac{4\beta P_{FE}}{(\alpha+\beta)\pi} \cdot [\sum_{n=1}^{\infty} \{\frac{1}{n}(\sin(\frac{n\pi\alpha}{\alpha+\beta}) - \sin(\frac{n\pi\beta}{\alpha+\beta}))\sin(\frac{2n\pi}{\alpha+\beta}x) \cdot e^{\frac{-2n\pi}{\alpha+\beta}z}\}]$$

$$+ \frac{8P_{FE}}{\pi^2} \cdot \sum_{n,m=1}^{\infty} \{\frac{1}{nm} \cdot (\sin(\frac{n\pi\alpha}{\alpha+\beta})\sin(\frac{m\pi\beta}{\alpha+\beta})) \cdot \frac{\sqrt{n^2+m^2}}{n} \cdot \cos(\frac{2n\pi}{\alpha+\beta}x)\cos(\frac{2m\pi}{\alpha+\beta}y) \cdot e^{\frac{-2\sqrt{n^2+m^2}\pi}{\alpha+\beta}z}\}$$

, (S7)

$$D_y(x,y,z) = D_x(y,x,z). \tag{S8}$$

Finally, with $\vec{D}(x,y,z)$, determined as described above, we analytically obtained the depolarization energy, $W_D$, domain wall energy, $W_W$, and interaction energy, $W_S$, between ferroelectric domains as a function of $E_S$ as below [5]:

$$W_D = \frac{1}{2}\int \vec{D} \cdot \vec{E} dV = \frac{1}{2\varepsilon}\int |\vec{D}|^2 dV$$

$$= \frac{V}{2\varepsilon}\{[\frac{P_{FE}((\alpha+\beta)^2 - 4\beta^2)}{(\alpha+\beta)^2}]^2 + 8 \cdot \frac{\beta^2}{(\alpha+\beta)} \cdot \frac{P_{FE}^2}{\pi^3} \cdot \frac{1}{t}\sum_{n=1}^{\infty}[\frac{1}{n^3}(\sin(\frac{n\pi\alpha}{\alpha+\beta}) - \sin(\frac{n\pi\beta}{\alpha+\beta}))^2]$$

$$+ (\frac{2P_{FE}}{\pi^2})^2 \cdot \frac{(\alpha+\beta)}{\pi} \cdot \frac{1}{t}\sum_{n,m=1}^{\infty}[(\frac{1}{nm})^2 \cdot \frac{1}{\sqrt{n^2+m^2}}(\sin(\frac{n\pi\beta}{\alpha+\beta})\sin(\frac{m\pi\alpha}{\alpha+\beta}) + \sin(\frac{n\pi\alpha}{\alpha+\beta})\sin(\frac{m\pi\beta}{\alpha+\beta}))^2]$$

$$+ (\frac{2P_{FE}}{\pi^2})^2 \cdot \frac{(\alpha+\beta)}{\pi} \cdot \frac{1}{t}\sum_{n,m=1}^{\infty}[(\frac{1}{nm})^2 \cdot \frac{1}{\sqrt{n^2+m^2}}(\sin(\frac{n\pi\alpha}{\alpha+\beta})\sin(\frac{m\pi\beta}{\alpha+\beta}))^2]\}$$

, (S9)

$$W_W = V[\frac{8\beta}{(\alpha+\beta)^2}\sigma], \tag{S10}$$

$$W_S = -\vec{P} \cdot \vec{E_S} = -V[\frac{(\alpha+\beta)^2 - 4\beta^2}{(\alpha+\beta)^2}]P_{FE}E_S, \tag{S11}$$

where $P_{FE}$, $\varepsilon$, $t$, $V$, and $\sigma$ are the absolute polarization, dielectric permittivity, film thickness, film volume, and domain wall energy per unit area, respectively.

We determined the stable configurations of the ferroelectric domains by finding $\alpha$ and $\beta$ that minimize the total energy $W_{tot} = W_D + W_W + W_S$ for a given $E_S$. We found that when $P_{FE}$ is small and $\varepsilon$ is large, $E_S$ can strongly affect the domain configuration, as shown in Figure S5. Furthermore, we obtained the temperature dependence of the ferroelectric domain configuration by considering the temperature-dependence of $P_{FE}$, $\varepsilon$, and $\sigma$ in Eqs. (S9)–(S11). For the temperature-dependent calculations, we extrapolated $P_{FE}$ and $\varepsilon$ at high temperatures from the mean field theory (i.e., $P_{FE} \propto (T_C - T)^{0.5}$ and $\varepsilon \propto (T_C - T)^{-1}$) using the reported experimental values of $P_{FE}$ and $\varepsilon$ at lower temperatures. We also assumed that $\sigma$ had a temperature dependence of $(P_{FE})^2/\varepsilon$.

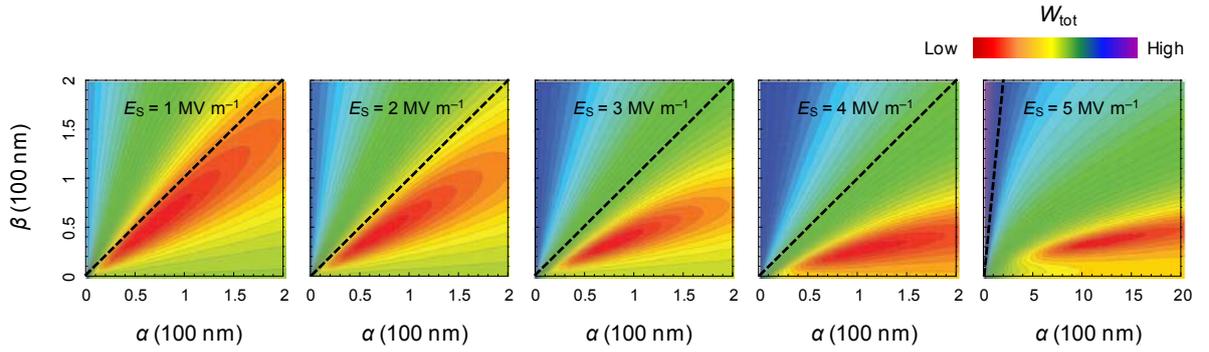

**Figure S5**. Dependence of $W_{tot}$ on $E_S$ for $P_{FE} = 0.5$ μC cm$^{-1}$ and $\varepsilon = 500\varepsilon_0$. Black dashed lines indicate the positions of $\alpha = \beta$.

Our calculated result (Fig. S6) shows that at $T \ll T_C$, the 180° polydomain (i.e., $(\alpha/\beta)^2 = 1$) is favorable. However, at temperatures near $T_C$, the 180° polydomain and the monodomain (i.e., $(\alpha/\beta)^2 \gg 1$) compete, depending on $E_S$. Specifically, as $E_S$ increases, the domain configuration changed from the polydomain to the monodomain. A larger $E_S$ will allow a wider temperature window where the ferroelectric monodomain is more favorable. These results indicate that the domain configuration can be tuned by the flexoelectric effect.

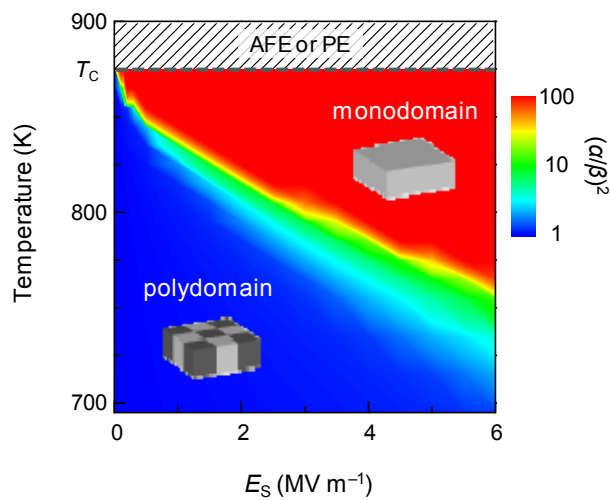

**Figure S6**. Variation of the domain occupation ratio, $(\alpha/\beta)^2$, as a function of $E_S$ and temperature, showing a strong dependence on $E_S$ at temperatures near $T_C$.

**References for Supplemental Material**